# Unconventional domain tessellations in moiré-of-moiré lattices

Daesung Park[1,2,11], Changwon Park[3,4,11], Kunihiro Yananose[3,11], Eunjung Ko[3], Byunghyun Kim[1], Rebecca Engelke[5], Xi Zhang[6], Konstantin Davydov[6], Matthew Green[6], Hyun-Mi Kim[7], Sang Hwa Park[2], Jae Heon Lee[2], Seul-Gi Kim[7], Hyeongkeun Kim[7], Kenji Watanabe[8], Takashi Taniguchi[9], Sang Mo Yang[2], Ke Wang[6], Philip Kim[5], Young-Woo Son[3*] and Hyobin Yoo[1,2,10*]

[1] Department of Materials Science and Engineering, Seoul National University, Seoul 08826, Republic of Korea

[2] Department of Physics, Sogang University, Seoul 04107, Republic of Korea

[3] Korea Institute for Advanced Study, Seoul 02455, Republic of Korea

[4] Department of Physics, Ewha Womans University, Seoul 03760, Republic of Korea

[5] Department of Physics, Harvard University, Cambridge, MA 02138, USA

[6] School of Physics and Astronomy, University of Minnesota, Minneapolis, MN, USA

[7] Korea Electronics Technology Institute, Seongnam 13509, Republic of Korea

[8] Research Center for Functional Materials, National Institute for Materials Science, 1-1 Namiki, Tsukuba 305-0044, Japan

[9] International Center for Functional Materials, National Institute for Materials Science, 1-1 Namiki, Tsukuba 305-0044, Japan

[10] Research Institute of Advanced Materials, Seoul National University, Seoul 08826, Republic of Korea

[11] These authors contributed equally.

**Corresponding authors**: Young-Woo Son (hand@kias.re.kr), Hyobin Yoo (hyobinyoo@snu.ac.kr)

## ABSTRACT

**Imposing incommensurable periodicity on the periodic atomic lattice can lead to complex structural phases consisting of locally periodic structure bounded by topological defects[1-8]. Twisted trilayer graphene (TTG) is an ideal material platform to study the interplay between different atomic periodicities, which can be tuned by twist angles between the layers, leading to moiré-of-moiré lattices[9-26]. Interlayer and intralayer interactions between two interfaces in TTG transform this moiré-of-moiré lattice into an intricate network of domain structures at small twist angles, which can harbor exotic electronic behaviors[9-26]. Here we report a complete structural phase diagram of TTG with atomic scale lattice reconstruction. Using transmission electron microscopy combined with a new interatomic potential simulation[27,28], we show several large-scale moiré lattices, including triangular, kagome, and a corner-shared hexagram-shaped domain pattern. Each domain is bounded by a two-dimensional network of domain wall lattices. In the limit of small twist angles, two competing structural orders-rhombohedral and Bernal stackings-with a slight energy difference, cause unconventional lattice reconstruction with spontaneous symmetry breaking and nematic instability, highlighting the importance of long-range interlayer interactions across entire van der Waals layers. The diverse tessellation of distinct domains, whose topological network can be tuned by the adjustment of the twist angles, establishes TTG as a platform for exploring the interplay between emerging quantum properties and controllable nontrivial lattices.**

## Main Text

Interfaces formed by joining two van der Waals (vdW) crystals has offered a new route to engineer interfacial electronic states by moiré superlattice[7,29-33]. At low twist angles, the atomic scale lattice reconstruction becomes appreciable within a superlattice domain, impacting the quasiperiodic electronic structures in the moiré superlattice. In twisted bilayer graphene (TBG), this lattice reconstruction effect can realize anomalous electronic transport and the strongly correlated behavior[7,29-32]. As the number of quasiperiodic interfaces increases in multilayer-twisted systems, extensive lattice reconstruction leads to the interplay between moiré superlattices formed within individual interfaces. Revealing the resulting hierarchical quasiperiodic structures modulated by the lattice reconstruction is essential for elucidating the unusual electronic structures recently reported in multilayer twisted graphene[9-26,34].

Twisted trilayer graphene (TTG) provides a simplified platform where the hierarchical structures are evident. The interference pattern between two moiré lattices, such as those between the bottom and middle layers and between the middle and top layers, generates moiré-of-moiré lattices governing electronic properties[9-26]. If the reconstruction strength is insufficient, the hierarchical structure remains incommensurate, exhibiting quasicrystal physics[21,22]. Recent studies[9-26] on various pairs of twist angles in TTG indicates the lattice reconstruction modulates the quasiperiodic moiré-of-moiré lattices. However, these studies have focused on twist angles typically exceeding 1.0°, where the reconstruction strength is insufficient to reveal the unique reconstruction characteristics in TTG.

We emphasize that TTG presents Bernal and rhombohedral stacking orders competing with slight but critical energy difference[35-38] unlike TBG. Due to the presence of two disparate low energy stackings in TTG, lattice reconstruction pattern can exhibit qualitatively different structural orderings upon tuning the reconstruction strength, providing a new pathway to engineer domain structures. This effect becomes more pronounced with the stronger lattice reconstruction, given the minute energy difference between Bernal and rhombohedral stackings. In our work, we establish an optimal platform to explore the unique atomic reconstructions in TTG by controlling the twist angles near 0.1°, an order of magnitude smaller than previous works, to promote stronger reconstructions. Recognizing the presence of various competing stacking orders in TTG[35-38], we investigate the impact of the long-range interactions across entire layers on the restructuring behavior, going beyond the typical emphasis on interactions limited to adjacent layers[39].

## Atomic reconstruction in TTG

The structure of TTG was investigated by exploiting electron diffraction and dark field (DF) imaging in transmission electron microscopy (TEM). We first fabricate the TTG specimen by controlling the two independent twist angles of $\theta_{12}$ and $\theta_{23}$ (Fig. 1a), employing the cut-and-stack method reported previously (see Method and Supplementary Section 1 for details)[10,40]. Figure 1b shows an optical microscopy image of the marginally TTG (MTTG) covered with the h-BN layers. As shown in the selected area electron diffraction (SAED) pattern in Fig. 1c, one cannot distinguish multiple sets of Bragg peaks of the graphene layers, suggesting that the TTG specimen was successfully fabricated with the small enough twist angles (see Supplementary Section 2 for details on twist angle determination).

DF TEM imaging reveals the emergence of tessellated commensurate domains along with their respective stacking orders resulting from atomic reconstruction. Observation of $g = 10\bar{1}0$ DF TEM image (Fig. 1d) and its variations with the sample's tilt angle (Fig. 1e) enables stacking order identification within the commensurate domains (See Method and Extended Data Fig. 1 for detailed diffraction analysis, and Supplementary Section 3 for atomic scale image analysis). Similarly, composite color DF TEM image obtained from the three sets of $2^{nd}$ order Bragg peaks (Fig. 1f) reveals the configuration of domain walls and elucidates their corresponding lattice shift vectors (See Method, Extended Data Fig. 2 and Supplementary Section 4 for details).

**Spontaneous symmetry breaking in atomic reconstruction**

Our DF TEM investigation on the MTTG specimen with $\theta_{12} = 0°$ and $\theta_{23} \cong 0.06°$ (Fig. 1d) reveals a regular arrangement of kagome-like domain lattice. It consists of six corner-sharing triangular domains surrounding inner hexagonal domains. These triangular domains alternate between rhombohedral ABC and ACB stackings (black and white regions in the inset of Fig. 1d) with the corresponding atomic models shown in Fig. 1g. Additionally, we identified a one-dimensional (1-D) array of domain walls (dark lines in Fig. 1d and blue lines in Fig. 1f) within each inner hexagonal domain, distinctly separating two different Bernal ABA and ACA stackings (grey regions in the inset of Fig. 1d with the corresponding atomic arrangements shown in Fig. 1g). Presence of 1-D array of domain walls inside the hexagonal domains reduces the original rotation symmetry of the MTTG, indicating the spontaneous symmetry-breaking (SSB) characteristics of the atomic reconstruction process.

We investigate the impact of the twist angle variations on the symmetry-breaking atomic reconstruction behavior in MTTG. Despite achieving precise manipulation over the twist angle through the cut-and-stack method[10,40], inherent variations in the twist angle persist due to limitations in control. In the MTTG specimen fabricated with $\theta_{12} = 0°$ and $\theta_{23} \cong 0.1°$, we observe a gradual change in the size of the moiré superlattice from left to right, with $\theta_{23}$ decreasing from 0.14° to 0.04° while $\theta_{12}$ remained at 0° (Fig. 2a). We observe a sharp structural phase transition from triangular domain lattice (left) to the kagome-like domain lattice (right) occurring at $\theta_{23} \cong 0.05°$, with its boundary marked by the black triangles (Figs. 2a and b). Further analysis reveals that alternating Bernal and rhombohedral (grey and white colored regions in Fig. 2a) stacking orders are formed in the triangular commensurate domains, exhibiting what we term 'a colored triangular domain lattice'. The transition from this lattice to the kagome-like domain lattice abruptly increases the prevalence of Bernal stacked regions from approximately 50% to 75%, implying the SSB in the reconstruction is related to the system favoring Bernal stacking over rhombohedral stacking.

**Origins of structural phase transition**

To understand the experimentally observed structural phase transition, we performed atomic force relaxations of a moiré unit cell using a newly developed interatomic potentials method[27,28] (see Method) to investigate the energetics of competing reconstruction patterns. Relative energies of the colored triangular and the kagome-like domain lattices (blue triangles and red stars in Fig. 2c) with respect to the Bernal stacked trilayer are plotted with varying $\theta_{23}$ while keeping $\theta_{12}$ at 0°. Simulated bright field (BF) TEM image (see Method for TEM image simulation) from the relaxed structures clearly shows alternations between Bernal and rhombohedral stacking orders for the colored triangular lattice (Fig. 2d) and six-corner sharing triangular rhombohedral stacking domains with the inner hexagonal Bernal stacking domains divided by 1-D domain walls (Fig. 2e), reproducing both experimental observations shown in Figs. 2a and b.

We note that the energies plotted for the two different lattice reconstructions increase linearly as the angle increases. This linear behavior can be captured by considering the competition between energy gain by increasing the low energy stacking area and simultaneous energy cost for domain wall formation. For small twist angles, the energy can be written as $E(\theta_{23}) = \frac{A_B(\theta_{23})E_B + A_R(\theta_{23})E_R}{A_B(\theta_{23}) + A_R(\theta_{23})} + \gamma|\theta_{23}|$ where $A_{B(R)}(\theta_{23})$ and $E_{B(R)}$ denote the angle-dependent area of Bernal (rhombohedral) stacked regions and their energies per unit area, respectively. We note that the first term in $E(\theta_{23})$ does not change with the angle unless the areal ratio between Bernal and rhombohedral stacking regions is altered with the angle and that the linear dependence on $\theta_{23}$ in the second term reflects the linearly increasing area-to-boundary ratio for each domain (See its derivation in Supplementary Section 5). From the slopes of the energies plotted for the two relaxed structures, the constants $\gamma$ for the colored triangular and kagome domains can be estimated to be 621 and 1013 meV per area of graphene unit cell, respectively. Then, we can deduce that the phase transitions between the two domain lattices should occur when the two energy lines (blue and red lines in Fig. 2c) intersect at $\theta_k \cong 0.025°$, which is consistent with the experimental observation of phase transition at $\theta_{23} \cong 0.05°$ (Fig. 2a, b, see Supplementary Section 2 and 6 for details on determining $\theta_k$). We note that the phase transition causes a discrete jump in the areal fraction of Bernal stacking, $\Delta A_B(\theta) = \frac{A_B(\theta)}{A_B(\theta)+A_R(\theta)}$, shifting from 50% to 75%.

Moreover, our theoretical investigation indicates that as the angle $\theta_{23}$ decreases further below 0.01°, an alternative reconstruction pattern or warping effect of domain boundaries emerge so that the whole area approaches to be the lowest energy configuration of Bernal stacking (See Supplementary Section 5 for details). As shown in Fig. 2c, the linear dependence of total energy does not hold if we consider distortion or warping of domain walls (See Supplementary Fig. S10 for pattern shapes). Therefore, for extremely small twist angles, two or three energetically degenerate domain patterns are possible.

Atomic reconstruction in the moiré systems involves rearrangement of atoms to maximize the area of lowest energy stacking configuration at the cost of domain wall formation[5,7,8]. In TBG, for instance, the energy differences between Bernal (AB) stacking and unstable AA stacking are order of 10 meV/atom[35-38], and thus upon twisting two layers below ~1°, entire region of the TBG specimen is dominated by Bernal stacked domains with triangular networks of domain walls[5,7,8]. With further decreasing angle, no additional structural transition occurs in TBG because no other competing order exists. For trilayer graphene, however, there are two low-energy stacking orders, Bernal and rhombohedral types (Fig. 1g). Though their energy difference is on the order of 0.1 meV/atom, this minute energy difference turns out to be crucial for the formation of various periodic domain structures in MTTG. Alongside well-known factors such as twist angles and local strains[5,7,8], we find that the most decisive factor to determine moiré structures in MTTG is a small difference between total energies for Bernal and rhombohedral stacking orders. This emphasizes that a simple moiré-of-moiré framework is insufficient and that the long-range interlayer interactions across the entire layers should be considered properly.

We further investigate the driving mechanism for the structural phase transitions by inspecting local atomic displacement patterns. Figures 2f and g are coarse-grained displacements of the colored-triangular and kagome domain lattices, respectively. The left, middle and right panels correspond to the displacement maps for the top (layer number 3), middle (2) and bottom (1) layers, respectively. With varying $\theta_{23}$ and keeping $\theta_{12} = 0$, only the top layer is twisted with respect to the others. As shown in Fig. 2f, the local displacement vectors near the vertices of triangular domains form same helical pattern for bottom and middle layer while it is reversed for the top layer with the twirling domain boundaries[41]. We also note that the local strains at the centers of triangular domains are negligible and that the bottom layer experiences less strains than the others. These strain patterns are similar with those shown in TBG[5,7,8]. However, unlike TBG, the adjacent triangular domains in Fig. 2d have no choice but to form different stackings with slightly different energies, thereby distinguishing the colored triangular domains in MTTG from simple triangular lattice in TBG.

Upon decreasing $\theta_{23}$ further, the atoms on the bottom layer overcome the local strain constraint imposed by the twisted top layer and are further relaxed to reduce the area of rhombohedral domain by moving atoms along a specific direction (Fig. 2g). Such reconstruction results in uniaxial strain fields or SSB from the three-fold rotational symmetry to two-fold one. As a result of this reconstruction, a kagome-like domain lattice with the nematic domain walls can be formed. This nematic boundary has three choices along three different diagonal directions of hexagonal domain. We observe two instances of the kagome-like domain lattice exhibiting distinct directions for the nematic ordering in the different regions of the same specimen (Figs. 2h and i, see Extended Data Fig. 3 for lower magnification image) highlighting their SSB nature. Note that the orientations of the nematic domain walls observed in Figs. 2h and i are rotated by 60° relative to each other.

**Complete structural phase diagram of TTG**

MTTG presents a fertile ground for diverse domain patterns with different symmetries due to the presence of the almost energetically equivalent yet different stacking orders. To investigate how the moiré superlattice domains manifest while independently varying the two twist angles $\theta_{12}$ and $\theta_{23}$, we performed comparative analysis between TEM imaging and simulations. A structural phase diagram of MTTG is presented in Fig. 3a. Our TEM observations are consistent with the simulated structures. Figs. 3b and 3d show experimental and simulated TEM domain contrast images, while Fig. 3c and e display the corresponding domain boundary contrast images in (see Extended Data Fig. 4 for the raw data). With the convention of twist angles defined in Fig. 1a, half of the first and second quadrants in $\theta_{12}$-$\theta_{23}$ space are enough to represent all possible twisting geometries (See discussions in Supplementary Section 7). The first quadrant corresponds to helical stacking while the second one to alternated stacking. The color scale in the diagram denotes the angle-dependent areal fraction of Bernal stacked region, $\Delta A_B(\theta) = \frac{A_B(\theta)}{A_B(\theta)+A_R(\theta)}$. The approximated expressions for energies of the various competing phases with the twist angles below 0.1° can be found in Supplementary Section 8.

In the TTG with alternated twist angles of the same magnitude ($\theta_{12} + \theta_{23} = 0$, marked with line 1 in Fig. 3a), 'a simple triangular domain lattice' emerges, where Bernal-stacked domains of ABA and ACA alternate with each other (panels (i) of Fig. 3b-e), similar to TBG. Away from the line of $\theta_{12} + \theta_{23} = 0$, slight mismatch between the two twist angles generates moiré-of-moiré lattice. Our observation on TTG specimen with $\theta_{12} = -0.3°$ and $\theta_{23} = 0.4°$, where the twist-angle combination deviates from $\theta_{12} + \theta_{23} = 0$ shows the simple triangular domain lattice surrounded by the hexagonal moiré-of-moiré domain wall network (panel (ii) of Fig. 3b-e). In the case of incommensurate twist-angle combinations as in our observation, atomic reconstruction occurs at the two interfaces, forming locally commensurate lattice bound by the discommensurate boundaries[2,21]. Depending on the details for morphologies of domain walls and twist angles, the resulting moiré-of-moiré domain wall network can be hexagonal (panels (ii) of Fig. 3b-e) or triangular (see Supplementary Section 9). Consequently, maximum areal fraction of Bernal stacking, $\Delta A_B = 1$, is maintained for a wide range of twist angles as noted by the yellow-colored region in Fig. 3a.

We note that there are discrete regions in the phase diagram where the combinations of the two twist angles satisfy the commensurate conditions (see Supplementary Section 10 for details). For instance, along the line of $2\theta_{12} + \theta_{23} = 0$ (marked with line 2 in Fig. 3a) where the ratio between the two moiré lengths becomes a rational number of 1/2, fully commensurate double colored triangular domain lattice is demonstrated (panels (iii) of Fig. 3b-e). In this configuration, the areal fraction of the Bernal-stacked region reduces to $\Delta A_B = 3/4$. Small deviation from the commensurate condition of $2\theta_{12} + \theta_{23} = 0$ forms larger-scale moiré-of-moiré domain walls with the corresponding areal fraction of the Bernal stacking, $\Delta A_B = 3/4$, preserved for a finite range of twist angle combinations (Fig. 3a) as it does similarly near $\theta_{12} + \theta_{23} = 0$. Our theoretical investigation suggests additional commensurate angle combinations (see Supplementary Section 10 for details), including one noted by the dashed line of $3\theta_{12} + \theta_{23} = 0$ (marked with line 3) surrounded by the areal fraction of Bernal stacking $\Delta A_B = 2/3$.

As $|\theta_{12}|$ decreases at a fixed $\theta_{23} > \theta_k$, we find that the normalized ratio of Bernal type stacking $\Delta A_B$ decreases stepwise from 1 to 3/4, to 2/3 and eventually to 1/2 at $\theta_{12} \cong 0$ (marked with line 4 in Fig. 3a) as denoted by different colors in Fig. 3a (See Extended Data Fig. 5 for transition between the different phases). As both twist angles approach $0°$ where the stacking energy contribution eventually dominates, we find the areal fraction of Bernal stacking converges to 1 regardless of the pathway taken in the phase diagram. One such pathway involves a series of phase transitions or warping effects of boundaries as denoted by the gradual color changes in the phase diagram.

Lastly, in the first quadrant of the phase diagram, as $\theta_{12}$ increases to approach the line of $2\theta_{12} - \theta_{23} = 0$ (marked with line 5) and $\theta_{12} - \theta_{23} = 0$ (marked with line 6) at a fixed $\theta_{23}$, the relaxed domain structures follow a similar physics with those near the lines of $2\theta_{12} + \theta_{23} = 0$ and $3\theta_{12} + \theta_{23} = 0$ on the second quadrant: two moiré lattices readjust themselves to form locally commensurate moiré domain structures with moiré-of-moiré domain walls. However, for the helical stacking cases, the two moiré structures are inverted with respect to each other (see Supplementary Section 10) so that the resulting super-domains show complicated colored triangular lattices corresponding to different stacking orders. We also note that they can exhibit $\Delta A_B = 7/12$ differing from that of the alternated stacking cases. Notably, along $\theta_{12} - \theta_{23} = 0$, the relaxed domain shows a peculiar moiré superlattice with corner-shared hexagram-shaped units, still maintaining six-fold rotational symmetry (panels (vi) of Fig. 3b-e).

We emphasize that the total energy difference between Bernal and rhombohedral stackings ($E_B$ and $E_R$), or equivalently the long-range interlayer interaction, is the most critical factor to determine relaxed domain patterns. As shown in Extended Data Fig. 6, the difference between $E_B$ and $E_R$ varies depending on the interlayer interactions between the first and third layers. By artificially controlling these interactions, one can relax the structures under two different energetic conditions, i.e., $E_B < E_R$ and $E_B > E_R$. As shown in Extended Data Fig. 7, completely different relaxed results can be obtained for twist angles corresponding to $\theta_{12} + \theta_{23} = 0$ and $\theta_{12} - \theta_{23} = 0$, respectively, thus highlighting a critical role of the long-range interactions in forming nontrivial superlattices (see comprehensive modeling in Supplementary Section 11). We also note that considering the inherent twist angle inhomogeneity in TTG, unavoidable even with state-of-the-art technique for fabricating twisted layered systems[10,40,42], sharp boundaries between disparate SSB structures could be possible in most of the MTTG specimens as illustrated in Fig. 2a and Extended Data Fig. 5.

**Discussions and Conclusions**

Having provided a complete structural phase diagram of TTG with twist angles below 0.1°, now we investigate its consequences for electronic properties. With typical domain sizes around hundreds of nanometer scale observed in our work, electronic structures of TTG may be dominated by bulk electronic properties of domains and their boundaries. Under perpendicular electric fields, the rhombohedral stacked region should open a band gap while the Bernal stacked region maintains its metallic nature[43,44]. Thus, we note several phases shown in Fig. 3 can provide unique platforms where various topological edge states can be realized when subjected to applied gate potentials. In the case of kagome-like domain lattices, for instance, diverse electronic states manifest at different junctions between distinct stacking orders. We note that the node region formed between two rhombohedral stacked regions, i.e., between ABC and ACB regions (shown in Fig. 1d and (v) of Fig. 3b) can host topological boundary states. With the applied perpendicular electric fields, both domains are fully gapped with distinct valley Chern numbers so that the topological boundary states develop as in TBG[45-49] (See Supplementary Section 12). Another type of the domain boundaries formed between two Bernal stacked domains inside the hexagonal shaped regions, on the contrary, do not host in-gap boundary states as the two Bernal stacked domains maintain their metallic nature. The other type of domain boundary is formed between rhombohedral and Bernal stacked regions where the valley Chern number cannot be determined well owing to metallic Bernal stacking[46,50]. However, as shown by our electronic structure calculations (Supplementary Fig. S17), we find a resonant state between localized boundary states and metallic states on Bernal stacked domains, suggesting the well-defined conducting channels along the boundaries in kagome shaped domains.

Owing to disparate local stacking structures and notable boundary states between them as discussed above, the kagome-like domain lattice can host distinct electronic states (See Method for electronic structure calculation). Without perpendicular electric field, local charge density indeed reflects different stacking geometries such that rhombohedral stacking region has a higher density than Bernal stacked hexagonal area with clear nematic boundary and the increased charge density at vertices between triangular shaped regions (Fig. 4a). With perpendicular electric field, all rhombohedral regions are fully gapped and all states at the Fermi energy are localized only in the Bernal-stacked area as shown in Fig. 4b. We note that triangular domain lattice shown in Fig. 2d also can show a similar variation in local electronic structure under the electric field depending on stacking geometries[33] (See Supplementary Section 13 for details). These localized states also clearly show the nematic characteristics indicated by the well-separating boundaries between two different Bernal-stacked regions. More interestingly, vertices have still in-gap states (Fig. 4c) and as shown in Fig. 4d, the charge density near the band edge for the gapped rhombohedral state shows a peculiar array of dumbbell-shaped states at the triangular shaped regions connected by the vertex boundary states. From these states, we note that the reconstructed domain lattices in TTG could provide a platform where we envisage intriguing interplays between various nontrivial correlated states since spatially well-defined rhombohedral stacked regions are separated by boundary states. The area of rhombohedral-stacked regions observed in several phases in TTG seems to be large enough to maintain bulk electronic properties so that the superconducting[51] and ferromagnetic states[52] can be realized in each rhomboheral-stacked domain with appropriate doping and field controls. Then, we may consider an interesting interplay[53] between magnetic insulator and superconductor located side-by-side seamlessly across the domain walls in the reconstructed twisted layered systems.

## Figure legends

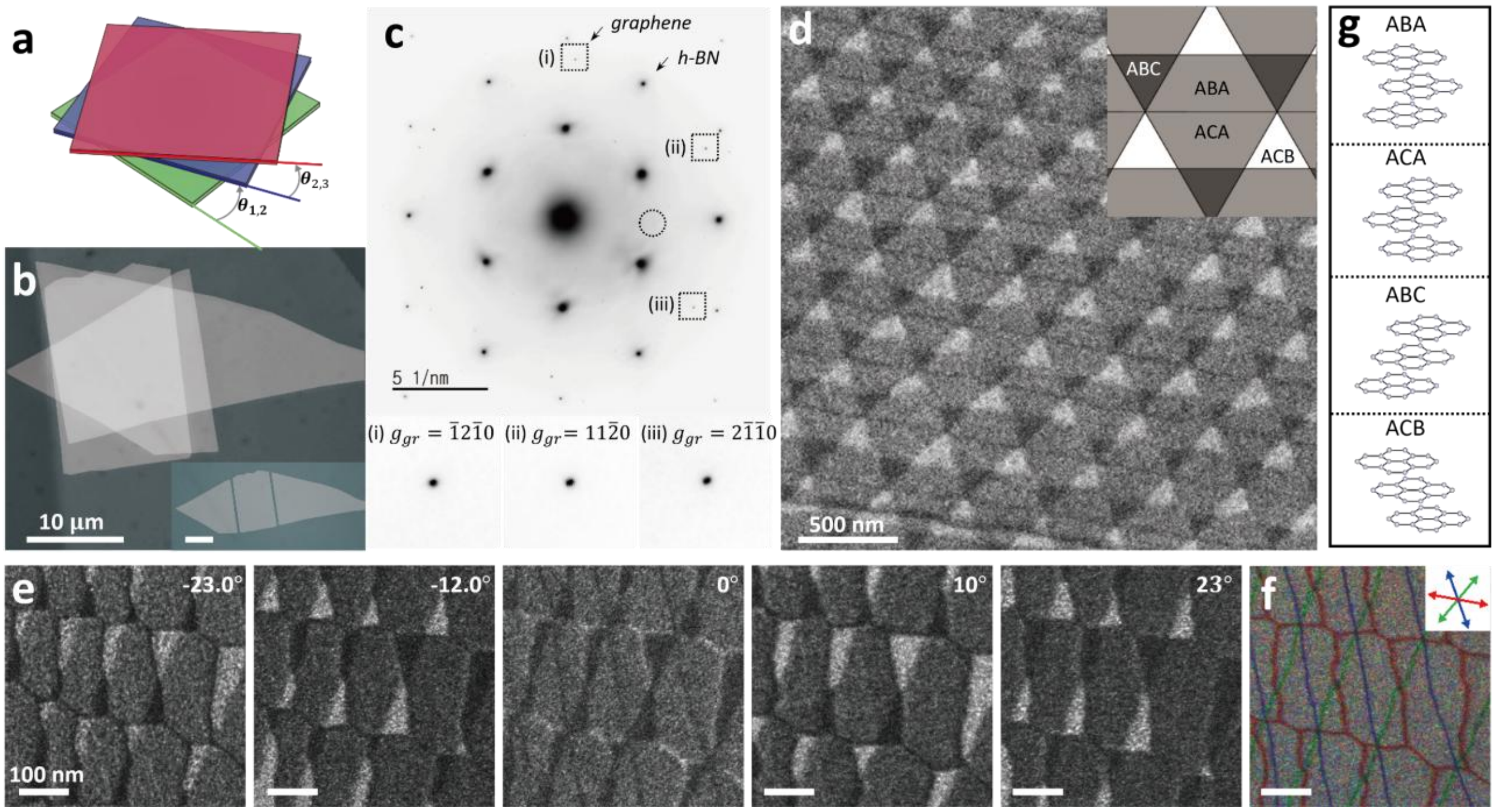

**Figure 1**. **Spontaneous-symmetry-breaking atomic reconstruction in TTG.** (a) Schematic illustration of the twisted trilayer graphene (TTG). Note that $\theta_{12}$ and $\theta_{23}$ can have both positive and negative values. The same graphene layers are colored differently to distinguish top, middle, and bottom layers. (b) Optical microscopy image of TTG. Monolayer graphene that was used to fabricate the TTG is shown in the inset. The monolayer graphene is segmented into three pieces before stacked to fabricate TTG. (c) Selected area electron diffraction pattern obtained from TTG covered with h-BN layers. $g = 10\bar{1}0$ Bragg peak is marked with a black dashed circle, and (i) $g = \bar{1}2\bar{1}0$, (ii) $g = 11\bar{2}0$, and (iii) $g = 2\bar{1}\bar{1}0$ Bragg peaks are marked with black dashed squares with their magnified views shown on the bottom. (d) Dark field (DF) transmission electron microscopy (TEM) image of TTG with $\theta_{12} = 0°$ and $\theta_{23} \cong 0.06°$. The DF TEM image was acquired by taking $g = 10\bar{1}0$ Bragg peak marked by the black dashed circle in (c). Schematic representation of the domain structure with the corresponding stacking orders is shown in the inset. (e) Tilt-series DF TEM images obtained from the TTG specimen with $\theta_{12} = 0°$ and $\theta_{23} \cong 0.07°$ by manipulating the tilt angle of the specimen in TEM (see details in Extended Data Fig. 1). (f) Composite color DF TEM image obtained from the three sets of 2nd order Bragg peaks $g = \bar{1}2\bar{1}0$, $g = 11\bar{2}0$, and $g = 2\bar{1}\bar{1}0$. Colored lines indicate the domain walls with the directions of the characteristic displacement vectors marked in the inset (see details in Extended Data Fig. 2). (g) Schematic illustration of Bernal stacked ABA and ACA configurations and rhombohedral stacked ABC and ACB configurations.

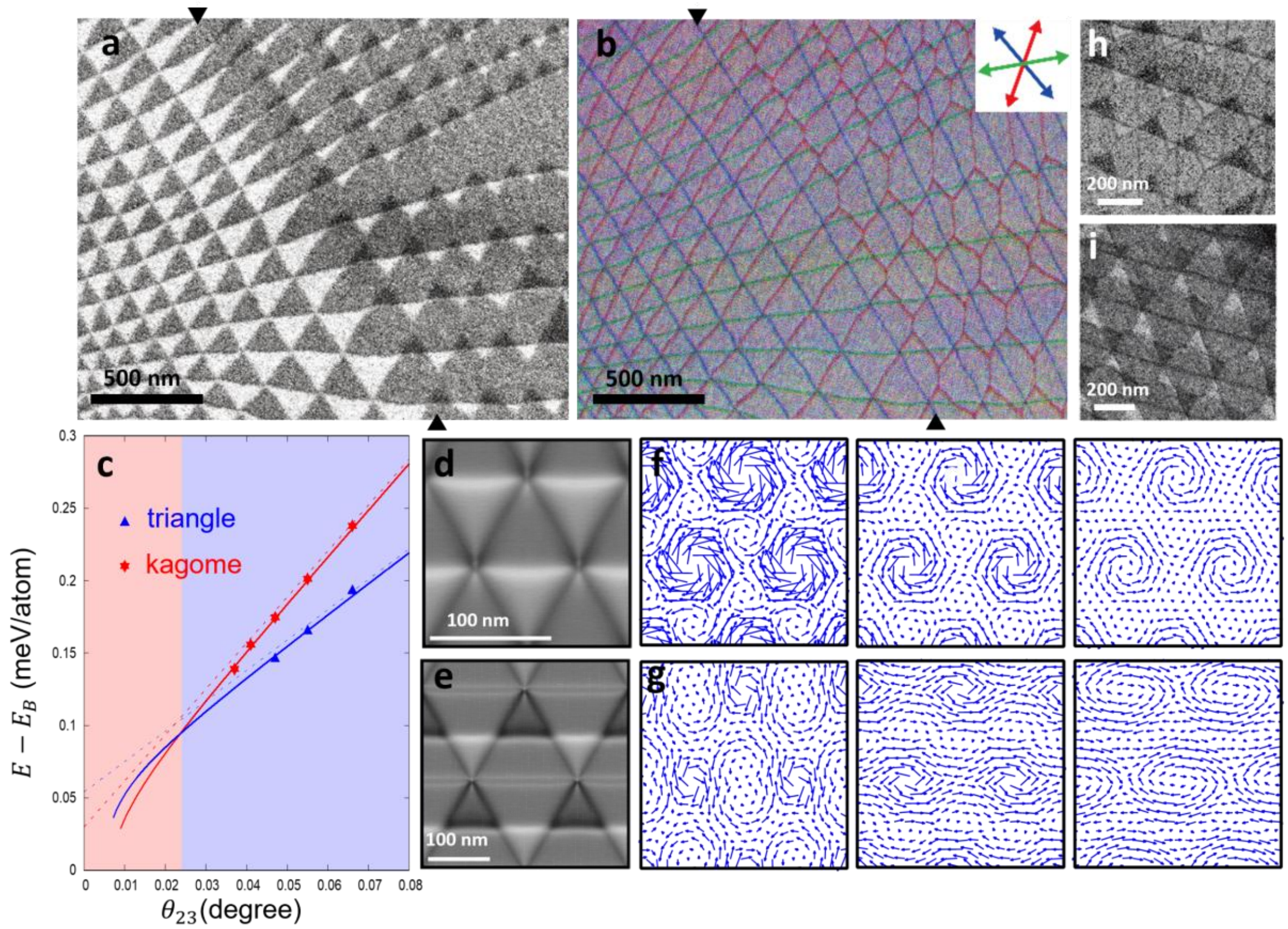

**Figure 2. Structural phase transition in TTG.** (a-b) DF TEM images of TTG with $\boldsymbol{\theta_{12} = 0}°$ and $\boldsymbol{\theta_{23} \cong 0.1}°$. While the angle $\boldsymbol{\theta_{12}}$ was kept at $\mathbf{0}°$, $\boldsymbol{\theta_{23}}$ gradually decreases from 0.14° (left) to 0.0$\mathbf{4}$° (right). $\boldsymbol{g = 10\bar{1}0}$ DF TEM image (a) and composite color DF TEM image ($\boldsymbol{g = \bar{1}2\bar{1}0}$, $\boldsymbol{g = 11\bar{2}0}$, and $\boldsymbol{g = 2\bar{1}\bar{1}0}$; see detailed description in Extended Data Fig. 2) (b) are obtained from the same region of the specimen to exhibit domain and domain boundary contrast, respectively. (c) The relative energies $\boldsymbol{E - E_B}$ of TTG with varying $\boldsymbol{\theta_{23}}$ while fixing $\boldsymbol{\theta_{12}} = 0$. $\boldsymbol{E_B}$ is the energy of the Bernal-stacked trilayer graphene. Calculated configurations are plotted with blue triangles and red stars for the colored triangular domain and kagome-like domain, respectively. Solid lines represent total energies with including warping effects of domain boundaries and dotted straight lines represent them without the effects. The background color variation points to the phase transition. (d-e) Simulated BF TEM image of the colored triangular (d) and kagome-like (e). Grey (black and white) area corresponds to Bernal (rhombohedral) stacking. (f-g) The coarse-grained atomic displacements of triangular (f) and kagome (g) domains. The left, middle and right panels correspond to the layer 3, 2 and 1, respectively. (h-i) Composite DF TEM images of TTG obtained from the region 1 with $\boldsymbol{\theta_{12} =}$ 0.0$\mathbf{6}$° and $\boldsymbol{\theta_{23} = 0}°$ (h), and the region 2 with $\boldsymbol{\theta_{12}} = 0°$ and $\boldsymbol{\theta_{23} = 0.05}°$ of the same specimen (see the larger field-of-view image in Extended Data Fig. 3). The composite DF TEM image was obtained by summing $\boldsymbol{g = 10\bar{1}0}, \boldsymbol{g =} \boldsymbol{\bar{1}2\bar{1}0}$, $\boldsymbol{g = 11\bar{2}0}$, and $\boldsymbol{g = 2\bar{1}\bar{1}0}$ DF images to visualize the domain and domain boundary contrast together (see details in Extended Data Fig. 3). Note that the orientations of nematic orders observed in regions 1 and 2 are rotated 60° with respect to each other.

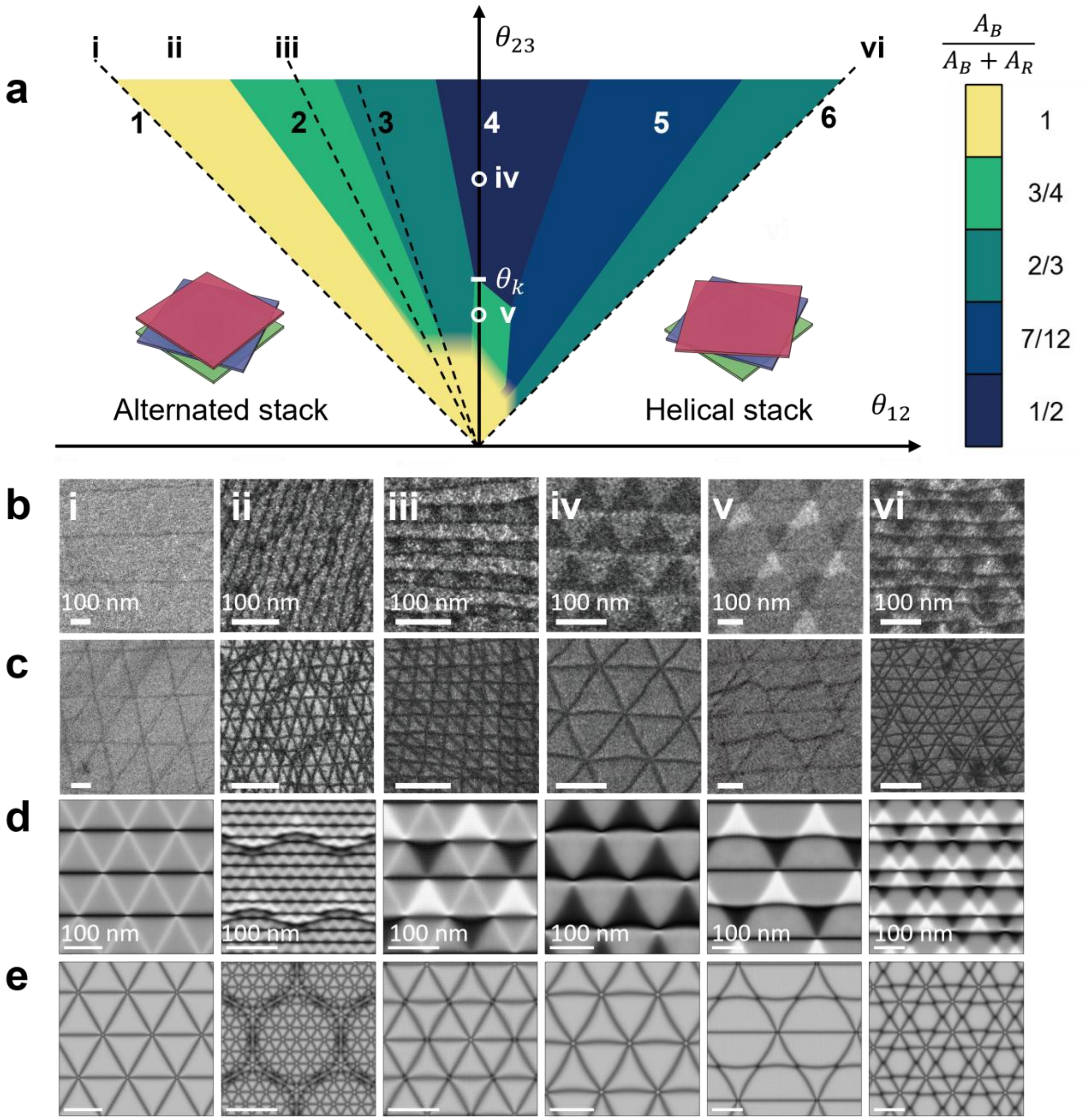

**Figure 3**. **Structural phase diagram of TTG.** (a) The phase diagram drawn as a function of $\boldsymbol{\theta_{12}}$ and $\boldsymbol{\theta_{23}}$**.** The color scale indicates the normalized areal fraction of Bernal-stacked region over the entire region, $\boldsymbol{\Delta A_B} = \frac{A_B}{A_B+A_R}$, where $\boldsymbol{A_B}$ and $\boldsymbol{A_R}$ denotes area of Bernal stacked region and rhombohedral stacked region, respectively. The lines marked with Arabic numerals represent a series of commensurate conditions which are surrounded by distinct phases noted by different colors: 1) $\boldsymbol{\theta_{12} + \theta_{23} = 0}$, 2) $\boldsymbol{2\theta_{12} + \theta_{23} = 0}$, 3) $3\boldsymbol{\theta_{12} + \theta_{23} = 0}$, 4) $\boldsymbol{\theta_{12} = 0}$, 5) $2\boldsymbol{\theta_{12} - \theta_{23} = 0}$, and 6) $\boldsymbol{\theta_{12} - \theta_{23} = 0}$. Experimentally observed phases are marked with roman numerals: i) simple triangular domain lattice, ii) simple triangular domain lattice bound by moiré-of-moiré boundaries, iii) double colored triangular domain lattice, iv) colored triangular domain lattice, v) kagome-like domain lattice, and vi) hexagram domain lattice. Near zero degrees for both $\boldsymbol{\theta_{12}}$ and $\boldsymbol{\theta_{23}}$, the transitions to $\boldsymbol{\Delta A_B = 1}$ become continuous owing to warping effects, as indicated by the gradual color changes. (b-c) DF TEM images obtained for the experimentally observed phases with the corresponding twist angles of i) $\boldsymbol{\theta_{12} = -0.06}°$, $\boldsymbol{\theta_{23} = 0.06}°$, ii) $\boldsymbol{\theta_{12} = -0.3}°$, $\boldsymbol{\theta_{23} = 0.4}°$, iii) $\boldsymbol{\theta_{12} = -0.2}°$, $\boldsymbol{\theta_{23} = 0.4}°$, iv) $\boldsymbol{\theta_{12} = 0}°$, $\boldsymbol{\theta_{23} = 0.09}°$, v) $\boldsymbol{\theta_{12} = 0}°$, $\boldsymbol{\theta_{23} = 0.05}°$, and vi) $\boldsymbol{\theta_{12} = 0.1}°$, $\boldsymbol{\theta_{23} = 0.1}°$. The 1st order

Bragg peak ($\boldsymbol{g} = \mathbf{10\bar{1}0}$) (b) and the sum of the 2nd order Bragg peaks ($\boldsymbol{g} = \mathbf{\bar{1}2\bar{1}0}$, $\boldsymbol{g} = \mathbf{11\bar{2}0}$, and $\boldsymbol{g} = \mathbf{2\bar{1}\bar{1}0}$) (c) are used to obtain DF TEM images to exhibit domain and domain boundary contrast, respectively (see Extended Data Fig. 4 for the raw data used to obtain domain boundary contrast images). (d-e) Simulated DF TEM images obtained for phases i)-vi). Simulated DF TEM images exploiting the 1st order Bragg peak (d) and the sum of 2nd order Bragg peak (e) reproduce the experimental observations shown in (b-c). All scale bars correspond to 100 nm.

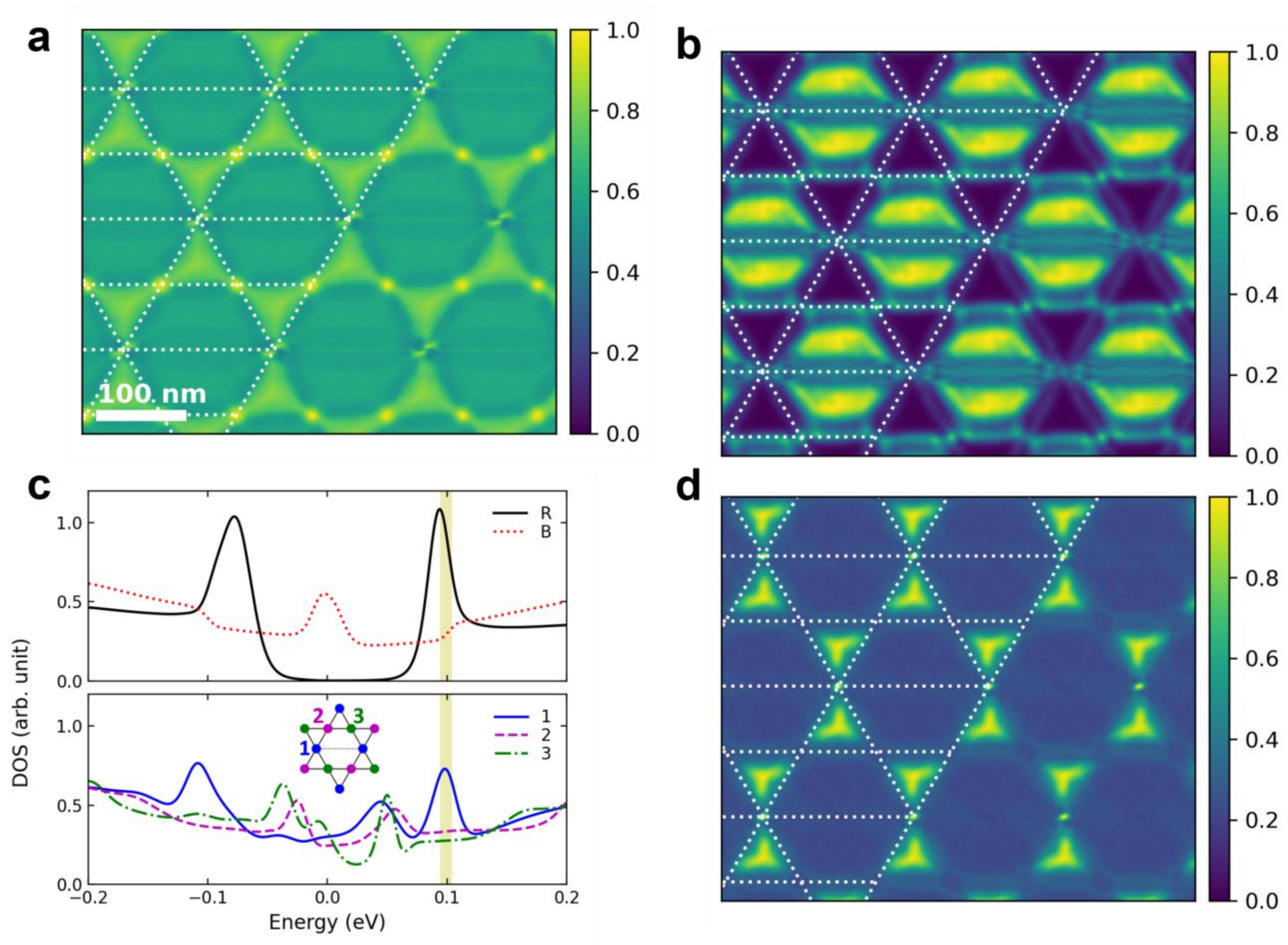

**Figure 4. Electronic structures of kagome-like domain pattern.** (a) Spatial density of states (DOS) for kagome-like domain lattice with $\theta_{12}$=0.08° and $\theta_{23}$=0° without perpendicular electric field. For DOS, we integrate the states of which energy is in between -50 and 0 meV. The Fermi energy is set to be zero. For all DOS's here, colormap is normalized to the maximum value, and the white dotted lines are drawn to highlight the domain walls as a visual guide. (b) Spatial DOS for kagome-like domain lattice near the Fermi level (from -10 meV to 0 meV) with the finite perpendicular electric field. (c) Energy-resolved DOS under the same electric field for rhombohedral (R) and Bernal (B) domain (top panel) and three different vertices (bottom panel, positions of each vertex indicated in the inset). To simulate a perpendicular field, we assign onsite energy shifts of +0.1, 0.0, and -0.1 eV for layer 1, 2, and 3, respectively. (d) Spatial DOS of which energy is denoted by the shaded range in c with the finite electric field.

## METHODS

**Sample fabrication.** Twisted trilayer graphene was fabricated by using ‘cut and stack’ method[10,40](See Supplementary Section 1 for details). We cut the monolayer graphene into three pieces by using contact mode of AFM to precisely control the twisted angle between each layer. Specifically, we conducted a line scan using a conductive AFM tip while applying the AC voltage. We used a worn AFM tip during the cutting process to widen the graphene flake’s pre-cut line width, which aids in aligning the h-BN with the graphene flake. The AFM conditions for cutting the monolayer graphene are as follows. i) The type of cantilever tip in AFM is a Pt/Cr-coated conductive tip (Multi75E-G, Budgetsensors). ii) The magnitude of tip loading force is about 60-90 nN. iii) The scan rate is 0.5 Hz. iv) The amplitude and frequency of AC voltage are 30 V and 100 kHz, respectively. The next step involves manufacturing a stamp that consists of a poly(Bisphenol A carbonate) coated on PDMS (polydimethylsiloxane). In a sequential process, we lift h-BN and three graphene pieces onto the stamp, carefully adjusting the twist angles. We used an engagement angle of approximately 3° between the stamp and each flake during sample fabrication process. To avoid potential structural rearrangement between the graphene and the h-BN layer, we deliberately offset the graphene from the top h-BN layer using the straight edges of the flakes. This offset angle is set to be greater than 3° to prevent closer alignment that could lead to such rearrangement[54]. Finally, we raise the temperature up to 180°C and transfer the twisted trilayer graphene fabricated on h-BN to a 20 nm thick SiN membrane chip. See Supplementary Section 1 for details in fabrication steps.

**TEM experiment.** We utilize the 80 kV and 200 kV field emission gun TEM (JEM 2100F, Jeol) equipped with CMOS camera (OneView, Gatan) to obtain selected area electron diffraction (SAED) patterns and dark field (DF) TEM images. Due to the limitations in angular resolution in SAED in reciprocal space and the spatial resolution of DF TEM imaging in real space, we rely on SAED patterns for resolving twist angles larger than 0.3° and on DF TEM images for resolving twist angles smaller than 0.3°. To estimate the twist angle from the moiré length $\lambda_m$, according to the relation $\lambda_m = a/(2\sin\frac{\theta}{2})$, where $a$ is the lattice constant of graphene, we measure and average the edge lengths of the moiré superlattice from DF TEM images obtained using the 2nd order Bragg peaks. The hexagonal unit cell resolved by DF TEM image is divided into two triangular half-unit cells, and the three edge lengths of each half-unit cell are averaged to estimate $\lambda_m$ for each half-unit cell. We note that the experimental measurements to estimate the twist angle were performed on moiré domains where strain effects are minimized, with distortions in the moiré unit cell limited to edge length variations of less than 30%. This allows for reliable comparison with theoretical estimation based on an ideal moiré superlattice (See Supplementary Section 2 and 6 for details).

For the DF TEM imaging, we utilized an aperture in the diffraction plane to isolate various types of Bragg peaks (denoted as '$g$'s). Depending on the Bragg peak one exploits, diverse spatial maps can be generated where each map provides distinct crystallographic information. We used an objective aperture with a diameter of 0.68 $nm^{-1}$ to select specific Bragg peaks. Electron dose rate used in our study was approximately $10^3$ $e^-/nm^2 \cdot s$, and the exposure time to take DF TEM image ranges from 10 to 60 seconds. A double tilt-holder was used to control the tilt angle of specimen in the microscope. Starting by aligning the specimen with [0001] zone axis, we capture a series of DF TEM images using the 1st order Bragg peak ($g = 10\bar{1}0$) while continuously changing the tilt angle of the specimen. By counting the amounts of electrons obtained from each domain region, we plot the intensity variation as illustrated in Extended Data Fig. 1c. These sinusoidal trends are generated by the interference of scattered electrons emanating from each layer in the TTG, containing three-dimensional structural information[5,55-58]. Analyzing the diffraction intensity variation as a function of the specimen tilt angle, the atomic registry of the domain formed within the TTG can be identified. Also, we utilize the TEM DF imaging with the 2nd order Bragg peaks to distinguish the lattice shift vector associated with each domain boundary[5,8]. After acquiring three different sets of TEM DF images using 2nd order Bragg peaks ($g = \bar{1}2\bar{1}0$, $g = 11\bar{2}0$, and $g = 2\bar{1}\bar{1}0$), we color-adjust the contrast of the three images with the red, green, and blue colors. By summing those three color-adjusted images, we obtain the composite color TEM DF images that visualizes domain boundaries (Fig. 1f, Fig. 2b, and Extended Data Fig. 2e) which are color coded according to the lattice shift vectors. See Extended Data Fig. 2 and Supplementary Section 4 for details. Spherical aberration ($c_s$) corrected annular dark field scanning transmission electron microscopy (Themis Z, Thermo Fisher Scientific) was employed to directly visualize the stacking orders within the domains. An acceleration voltage of 80 kV was used, along with a probe convergence angle of 27 mrad and collection angles of 42 mrad (inner) and 200 mrad (outer).

**TEM simulation.** For each optimized atomic structure, simulated TEM DF image is obtained from the diffraction intensity $I$ at pixel $\boldsymbol{r}$ as $I(\boldsymbol{r}) = \left|\sum_j \exp\left[i\left(\boldsymbol{r}-\boldsymbol{R}_j\right)\cdot(\boldsymbol{k}_O-\boldsymbol{k}_I)\right]\right|^2$ where $\boldsymbol{k}_I$ and $\boldsymbol{k}_O$ are wavevectors of incident and elastically scattered electron beam satisfying $|\boldsymbol{k}_I| = |\boldsymbol{k}_O|$, and $\boldsymbol{R}_j$ is a position vector of atom $j$. The summation is done for atoms in a cylinder which is centered at $\boldsymbol{r}$ with a radius 12 Å and tilted by the incident angle $\theta_{in}$. Because the relative in-plane displacements of atoms are very small within the cylinder, $I(\boldsymbol{r})$ is sharply peaked at $\boldsymbol{k}_O = \boldsymbol{k}_I + \boldsymbol{G}_{||} + \Delta k_z\hat{\boldsymbol{z}}$ where $\boldsymbol{G}_{||}$ is reciprocal vectors of a monolayer graphene and $\Delta k_z\hat{\boldsymbol{z}}$ is the change of wavevector along the out-of-plane direction of $\hat{\boldsymbol{z}}$. For $|\boldsymbol{k}_I| \gg \left|\boldsymbol{G}_{||}\right|$ which is the situation for usual TEM DF setup, the elastic scattering condition becomes $\boldsymbol{k}_I\cdot\boldsymbol{G}_{||} = -\boldsymbol{k}_I\cdot\Delta k_z\hat{\boldsymbol{z}}$. Specifically, when the in-plane component of $\boldsymbol{k}_I$ is parallel to $\boldsymbol{G}_{||}$, $\Delta k_z = \left|\boldsymbol{G}_{||}\right|\tan\theta_{in}$ and the diffraction intensity at $\boldsymbol{G}_{||}$ for an untwisted $n$-layer graphene becomes $I(\boldsymbol{r};\boldsymbol{G}_{||}) = \left|\sum_{k=1}^{n}\exp\left[i\boldsymbol{t}_k\cdot\boldsymbol{G}_{||} + d\left|\boldsymbol{G}_{||}\right|\tan\theta_{in}\right]\right|^2$ where $\boldsymbol{t}_k$ is overall translation of $k$-th layer and $d$ is the interlayer distance. Simulated TEM bright field (BF) images are obtained from the transmittance of electron beam, which is approximately calculated for each unitcell as the area of unitcell subtracted with the projected area of atomic spheres of radius 0.5 Å. Here, the direction of projection is given by $\theta_{in}$. For maximum contrast between Bernal- and rhombohedral-stacking area, $\theta_{in}$ is set to be $d\tan\theta_{in} = 0.81$ and 1.70 Å for TEM DF and BF simulations, respectively. For 2nd order Bragg peak simulations, $\theta_{in}$ is set to be zero for the intensities of Bernal (rhombohedral) stacking area and brightest domain boundary to be the same.

**Interatomic potentials for lattice relaxations.** Our simulation method is designed to reproduce *ab initio* calculation results in molecular dynamics for very large systems accurately and can describe complex potential landscapes for delicate structural phase transitions in various materials reliably[27,28]. To describe interlayer interactions, we add the pairwise interaction from Kolmogorov-Crespi (KC) potential[39]. Even though binding energies of various stacking geometries in TBG are well captured by the KC potential[39], we found that its naïve application to trilayer graphene results in an erroneous preference for rhombohedral stacking over Bernal one by 2.634 meV/nm$^2$. This is in contrast to experiment[35-37] and first-principles calculations[38] pointing to the Bernal stacking as the ground state of trilayer graphene. Therefore, we modify the KC potential to reflect small energetic gain of Bernal stacking over rhombohedral one by adding an additional long-ranged interlayer interaction between the topmost and bottommost layers in the trilayer.

Our interatomic potential is the sum of intra- and interlayer potential by treating them independently. Thanks to the enormous elastic constants of graphene, the maximum strain of twisted graphene layers is estimated to be smaller than 1% from our calculation so that we can safely neglect anharmonic effects. Also, by exploiting a fact that no bonds between carbon atoms break or wildly alter during the optimization, the computational cost can be further reduced and straightforwardly parallelized to relax about ten million atoms. The harmonic intralayer potential can be readily obtained from the phonon dispersion relation of monolayer graphene. For the reference phonon calculation, we have used density functional perturbation theory[59] with the generalized gradient approximation for exchange-correlation functional[60]. For interlayer potentials, we have used the pairwise interatomic potential developed by Kolmogorov and Crespi (KC potential)[39] that distinguishes energetic differences depending on the registry between two graphene layers. Since the corrugations in TTG is negligibly small, the KC potential between two carbon atoms on different layers connected by $\boldsymbol{r} = (x, y, z)$ can be written as $V(\boldsymbol{r}) = e^{-\lambda(r-z_0)}\left[C + 2f\left(r_{||}\right)\right] - A(r/z_0)^{-6}$ where $r = |\boldsymbol{r}|$, $r_{||}^2 = r^2 - z^2$ and $f\left(r_{||}\right) = e^{-\left(r_{||}/\delta\right)^2} \sum_{n=0} C_{2n}\left(r_{||}/\delta\right)^{2n}$. We use a set of parameters following a previous work[39], $\lambda = 3.629$ Å$^{-1}$, $z_0 = 3.34$ Å, $A = 10.238$, $\delta = 0.578$ Å, $C_0 = 15.71$, $C_2 = 12.29$, and $C_4 = 4.993$ and the potential is in unit of meV. With these equations applied to interlayer interactions between bottom and middle layers and to those between middle and top layers, respectively, the rhombohedral trilayer graphene is energetically favored over Bernal trilayer by 0.023 meV/atom or 2.634 meV/nm$^2$. This is because a tiny spurious alternating out-of-plane displacement (less than 0.1% of interlayer distance) for nearest interlayer carbon atoms denoted by red arrows in Extended Data Fig. 6a, b. To correct the ground state configuration to be Bernal type, we introduce a simple Gaussian potential between top and bottom layers, $V_{2nn}(\boldsymbol{r}) = -Ve^{-3(|r|^2 - z^2)} = -Ve^{-3r_{||}^2}$ as shown in Extended Data Fig. 6c. With $V_{2nn}(\boldsymbol{r})$, the difference between the energy of Bernal ($E_B$) and one of rhombohedral stackings ($E_R$) can be linearly tunable as shown in Extended Data Fig. 6d. We set $V = 0.5$ meV corresponding to $E_B - E_R = -0.06$ meV/atom to reproduce the experimental structural phase transition between triangle and kagome-shaped phases. We checked that details of correcting Gaussian potential hardly change energy configurations. For example, we confirmed that doubling the width of our Gaussian potential changes the total energy of kagome patterns less than 0.01 meV/atom.

**Estimation of domain wall energies.** There are two distinct domain walls in TTGs. One boundary is in between two Bernal stacked regions (type 1) and the other in between Bernal and rhombohedral stacking regions (type 2) as shown in Extended Data Figs. 8a and b, respectively. As shown in Extended Data Fig. 8c, the corrugations in the top and bottom layer along the type 1 domain wall are symmetric with respect to the flat middle layer. Unlike the symmetric type 1 domain wall, the corrugations along the type 2 in Extended Data Fig. 8d are asymmetric such that the corrugation on the top layer is induced from that of middle layer with much smaller amplitudes than that of type 1.

Simultaneously, the bottom layer distorts less than the corresponding type 1 wall. Since the domain wall energy mainly come from curvature of corrugations, we expect that the domain wall energy of type 1 could be larger than that of type 2. Because the energy of TTG at small angle linearly depends on the scaling of both twist angles as shown in Extended Data Fig. 8e, we can extract the energy of domain wall from the slope of the energy on the twist angle, Counting the total length of domain wall per moiré unitcell, the calculated wall energies are 137 and 84 meV/Å for type 1 and 2 domain walls, respectively. In the similar way, we can also compute moiré-of-moiré domain walls from atomistic simulations based on interatomic potentials.

**Electronic structure calculations.** We use the spinless tight-binding approximation with one $p_z$-orbital per each carbon atom. The matrix element between *i*-th and *j*-th atoms in the Hamiltonian matrix at a crystal moment $\mathbf{k}$ is given as $\mathrm{H}_{ij}(\mathbf{k}) = \sum_{\mathbf{R}} t_{ij\mathbf{R}} \exp(i\mathbf{k}\cdot\mathbf{R})$, where $\mathbf{R}$ is the lattice translation vector. For the inter- and intra-layer hoppings, we used $t_{ij\mathbf{R}} = t(r) = n^2 V_{pp\sigma}(r) + (1-n^2)V_{pp\pi}(r)$ following a previous work[61], where $r = |\mathbf{r}_{ij\mathbf{R}}| = |\mathbf{r}_j - \mathbf{r}_i + \mathbf{R}|$ is the distance between the atom $i$ and $j$ and $n = z/r$ is the *z*-direction cosine of $\mathbf{r}_{ij\mathbf{R}}$. Each exponentially decaying Slater-Koster parameter is given by $V_{pp\pi}(r) = -\gamma_0 \exp\,[q_\pi(1 - r/a_\pi)]$ and $V_{pp\sigma}(r) = \gamma_1 \exp\,[q_\sigma(1 - r/a_\sigma)]$ inside the cut-off radius $r_{\mathrm{cut}} = 5$ Å for the interlayer hopping and we only consider the nearest neighbor bonds for the intralayer hopping where $\gamma_0 = 2.7$ eV, $\gamma_1 = 0.48$ eV, $a_\pi = 1.418$ Å, $a_\sigma = 3.349$ Å, and $q_\pi/a_\pi = q_\sigma/a_\sigma = 2.218$ Å$^{-1}$. The applied electric field perpendicular to the layer is included by setting on-site energy difference between the layers, $+\Delta U = 0.1$ eV for the top layer, 0 eV for the middle layer, and $-\Delta U$ for the bottom layer, respectively. We consider a specific kagome-like domain lattice TTG with $\theta_{12} = 0.083°$ and $\theta_{23} = 0°$ that includes 2,887,206 carbon atoms in the moiré unit cell with its lattice constant of 1706.13 Å. We investigated the electronic structures near the Fermi level by using ARPACK eigenvalue-solver based on the implicitly restarted Lanczos method[62]. In addition, PETSc library is used as linear equation solver for the shift-invert mode[63]. Density of states (DOS) were calculated by gaussian broadening of energy eigenvalues with σ = 6 meV. We confirmed that the DOS is converged enough at 4×4 gamma-centered momentum space grid. In addition, site-projected DOS were evaluated from the weight of orbitals belonging to each site.

**Data Availability**

The data that support the findings of this study are presented in the paper, Extended Data, and Supplementary Information. Any other relevant data are available from the corresponding authors upon request.

**Methods-only references**

**Acknowledgements** D.P., B.K., and H.Y. were supported by the National Research Foundation of Korea (NRF) (Grant No. RS-2021-NR061606). H.Y. were supported by the National Research Foundation of Korea (NRF) (Grant No. RS-2021-NR060087, NRF-2022R1A4A1033562) and POSCO Science Fellowship of POSCO TJ Park Foundation. C.P. was supported by the new generation research program (CG079701) at Korea Institute for Advanced Study (KIAS). E.K. was supported by a KIAS individual grant (CG075002). K.Y. was supported by a KIAS individual grant (CG092501). Y-W.S. was supported by the KIAS individual Grant No. (CG031509). TEM work was supported by the Center for Materials Analysis at Research Institute of Advanced Materials (RIAM), Seoul National University, and the Center for Nano Materials at Sogang University. Computations were supported by the Center for Advanced Computation of KIAS. X.Z., K.D., M.G and K.W. were supported by NSF DMREF Award 1922165. H.-M.K., S.-G. Kim, and H. Kim acknowledges support from Material Innovation Leading Project through the National Research Foundation of Korea (NRF) funded by the Ministry of Science and ICT (2020M3H4A3081879). S.M.Y. was supported by the National Research Foundation of Korea (NRF) (Grant No. NRF-2023R1A2C1003047). P.K. acknowledges the support from DOE (DE-SC0012260) for sample preparation and ARO grant W911NF-21-2-0147 for analysis.

**Author Contributions** H.Y. conceived the experiments. D.P., B.K., R.E., X.Z., K.D., M.G., S.H.P., and J.H.L. fabricated samples. D.P., B.K., R.E., H.-M.K., S.-G.K., and H.K. performed TEM experiments. D.P., C.P., E.K., B.K., H.-M.K., S.M.Y., K.W., P.K., Y.-W.S., and H.Y. analyzed the data. C.P., E.K., K.Y., and Y.-W.S. performed theoretical analysis. K.W. and T.T. grew bulk hBN crystals. D.P., C.P., Y.-W.S., and H.Y. wrote the manuscript. All authors contributed to the overall scientific interpretation and edited the manuscript.

**Competing Interests** The authors declare no competing interests.

**Additional Information**

**Supplementary information** is available for this paper.

**Correspondence and requests for materials** should be addressed Y.-W.S. and H.Y.

**Reprints and permissions information** is available at www.nature.com/reprints.

**Extended Data legends**

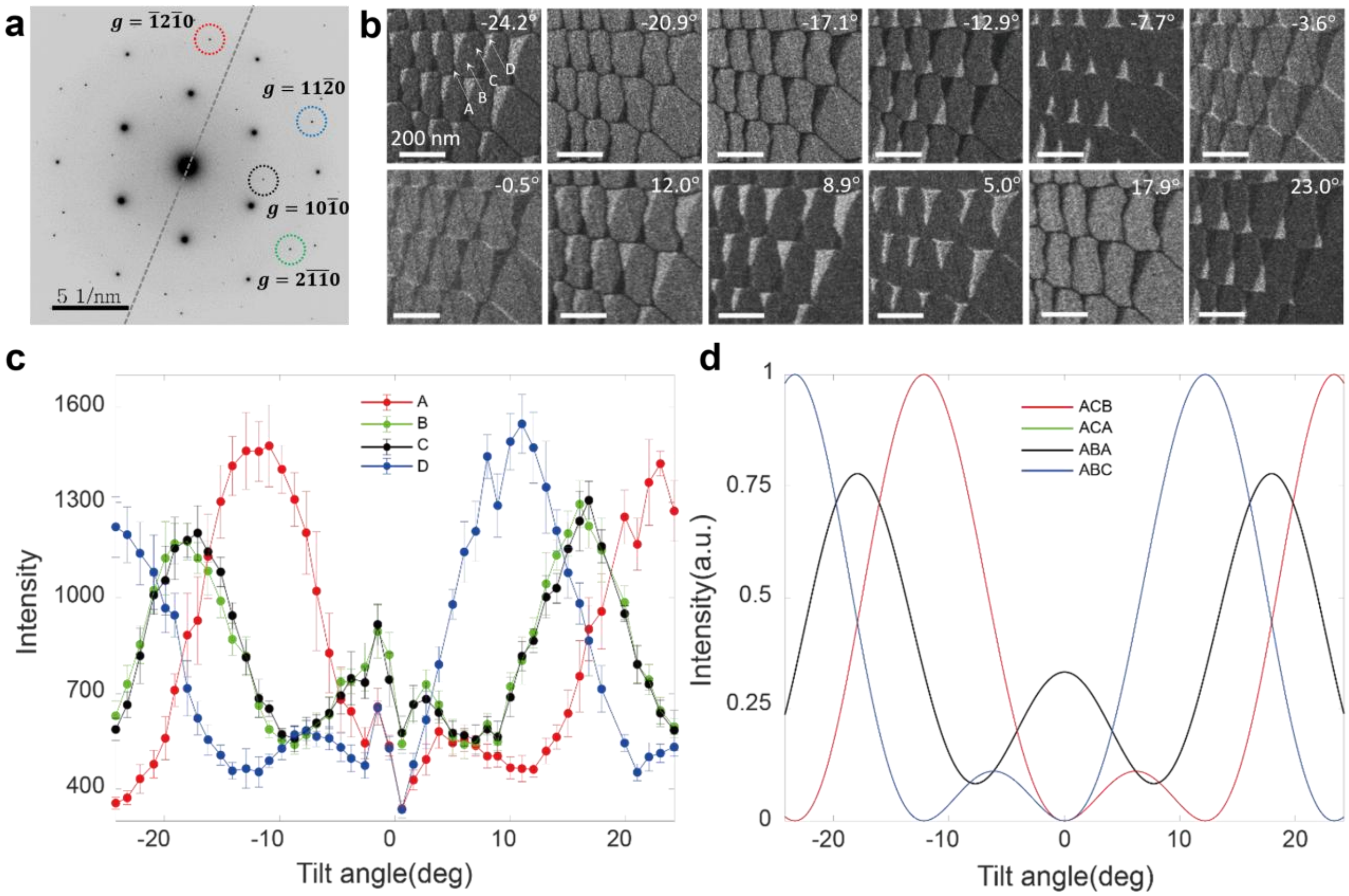

**Extended Data Fig. 1. Domain contrast image obtained by dark field (DF) transmission electron microscopy (TEM) analysis.** (a) Selected area electron diffraction (SAED) pattern obtained from twisted trilayer graphene (TTG). The 1st order Bragg peak ($g = 10\bar{1}0$) is marked with the black dashed circle, and the three sets of 2nd order Bragg peaks, $g = \bar{1}2\bar{1}0$, $g = 11\bar{2}0$, and $g = 2\bar{1}\bar{1}0$ are marked with red, blue, and green dashed circles, respectively. A grey dashed line is drawn to denote the tilt axis. (b) A series of DF TEM images obtained as a function of specimen tilt angle in the electron microscope. Four distinct domains are marked with A, B, C, and D. (c) Electron diffraction intensities experimentally measured from the four different domain regions as a function of tilt angle of the specimen. The red, green, black, and blue curves correspond to the diffraction intensities obtained from the domains marked with A, B, C, and D, respectively. The error bars represent the standard deviations of the intensities obtained from different domains with identical atomic configurations. (d) Simulated electron diffraction intensities obtained from different stacking orders in trilayer graphene. The red, green, black, and blue curves correspond to the simulated diffraction intensity obtained from ACB, ACA, ABA, and ABC stacking orders, respectively.

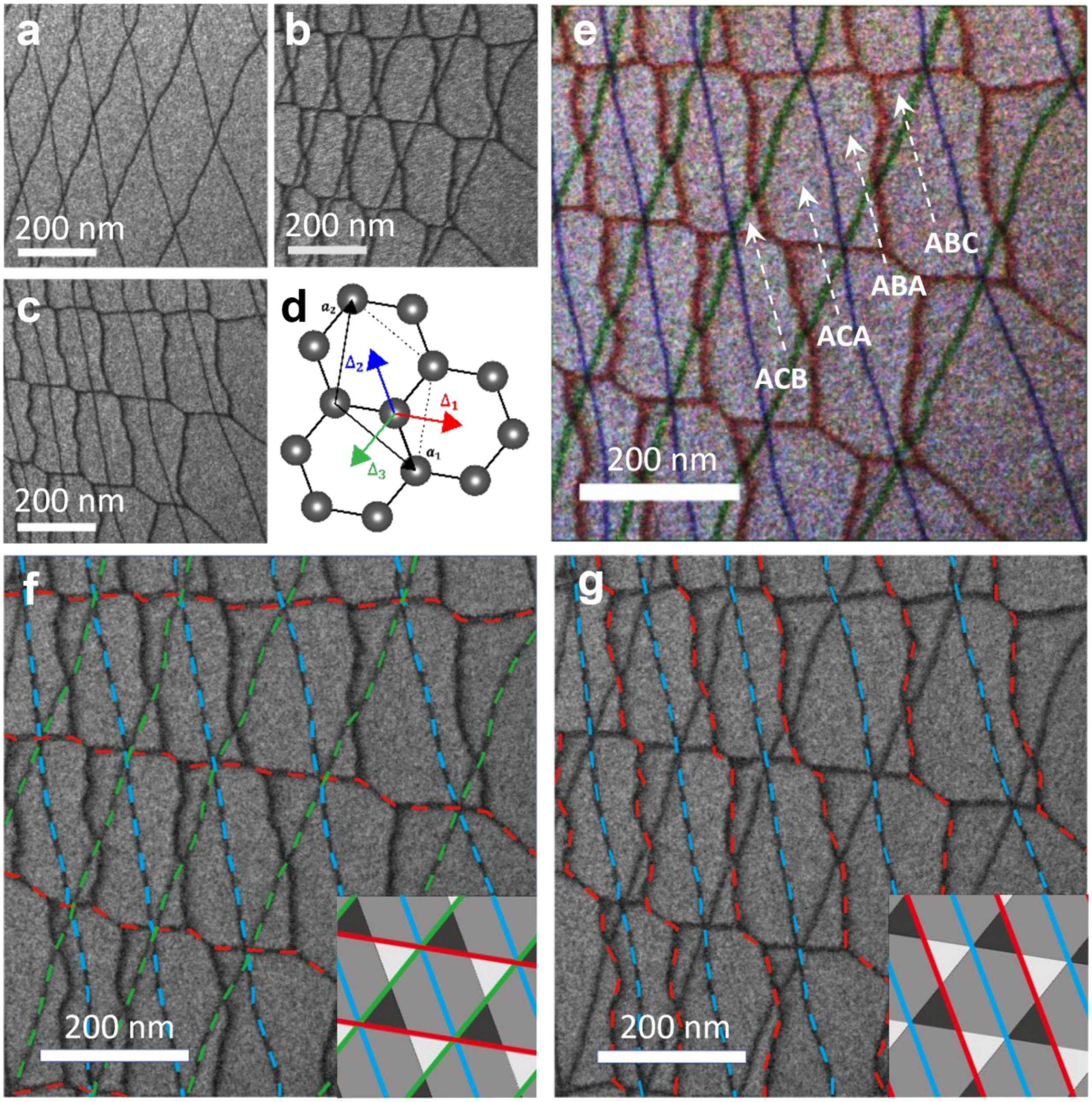

**Extended Data Fig. 2. DF TEM domain boundary contrast image obtained by DF TEM analysis.** (a-c) DF TEM images obtained from the 2nd order Bragg peaks. $g = \bar{1}2\bar{1}0$, $g = 11\bar{2}0$, and $g = 2\bar{1}\bar{1}0$ Bragg peaks are utilized to obtain the DF images shown in (a), (b), and (c), respectively. (d) Schematic image illustrating the displacement vectors $\Delta_i$ $(i = 1, 2, 3)$ associated with the domain boundaries. The displacement vectors $\Delta_i$ are drawn on top of the atomic structure to denote the directions and magnitudes of the displacements. (e) Composite color DF TEM image obtained from the three sets of DF images shown in Extended Data Fig. 2 a-c by summing them after color contrasting the individual images. As a result, the colored lines indicate the domain walls with the characteristic displacement vectors shown in (d). (f-g) Distinct domain boundary networks formed at the adjacent interfaces. Domain boundary network formed at the bottom interface (f) and the top interface (g) are drawn as colored dashed lines. When the bottom interface is twisted with finite angle ($\theta_{23} \neq 0°$), triangular domain wall network appears as a result of the lattice reconstruction as in (f). When the top interface exhibits the twist angle $\theta_{12} = 0°$, a parallel array of domain walls appears. The insets in (f) and (g) represent the schematic drawings for kagome-like domain lattices, where black and white regions represent the rhombohedral stacking orders, and the grey regions indicate the Bernal stacking orders. The domain walls are visualized with distinct colored lines, corresponding to the displacement vectors shown in (d)

with the same colors. In the twisted interface (f), the directions of each domain wall are mostly parallel with the displacement vector, indicating all the domain walls can be characterized with the shear type of displacements. In the untwisted interface (g), one type of domain walls (marked with red line in (g)) exhibits the displacement vector that has non-zero orthogonal component to the domain wall direction, revealing that the uniaxial displacement is incorporated along such domain walls.

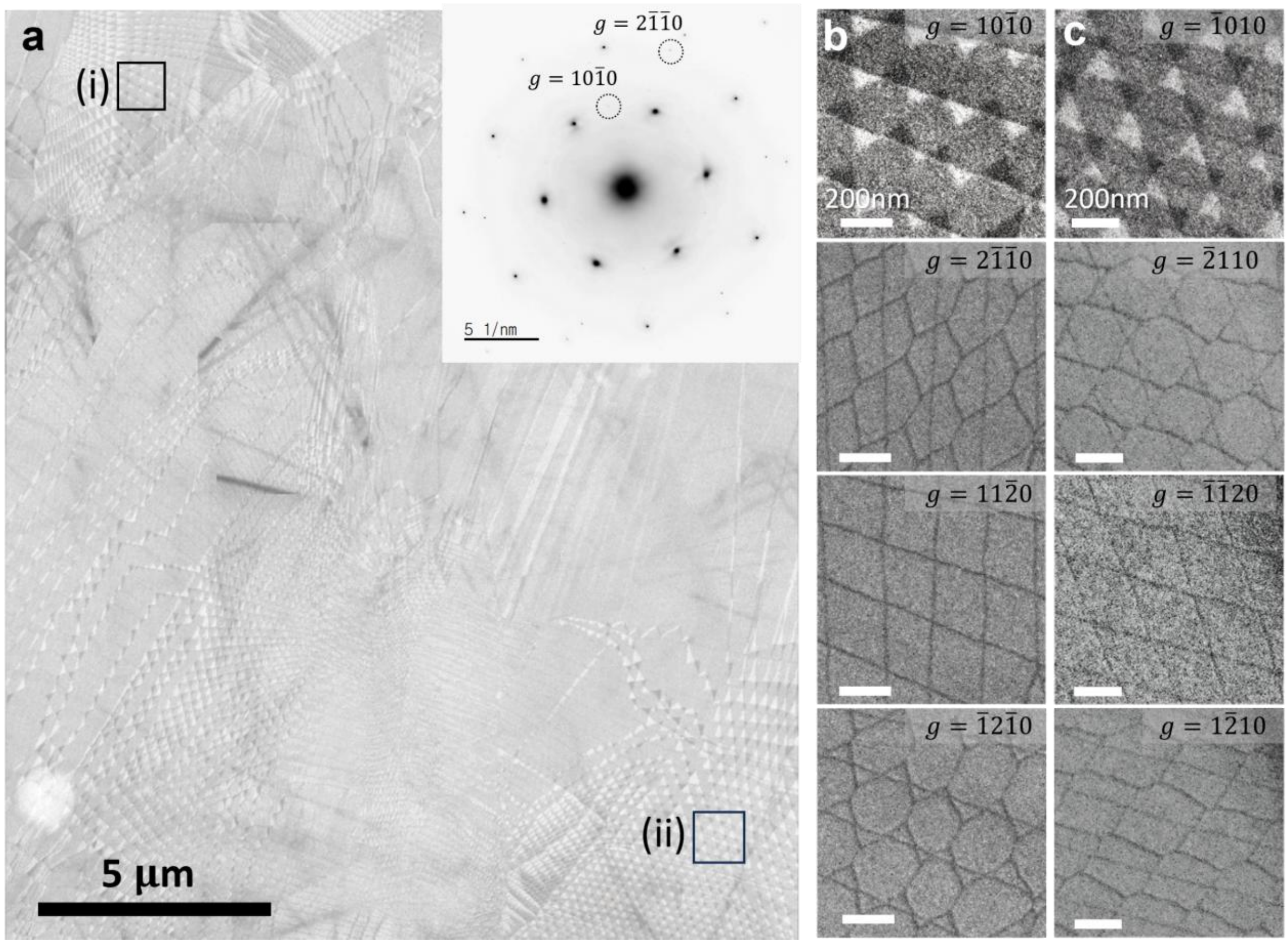

**Extended Data Fig. 3. Spontaneous symmetry breaking lattice reconstruction with distinct nematic orders.** (a) Low-magnification DF TEM image utilizing the 1st order Bragg peak to visualize the domain contrast. Inset shows the corresponding SAED pattern. (b) $g = 10\bar{1}0, g = 2\bar{1}\bar{1}0, g = 11\bar{2}0$, and $g = \bar{1}2\bar{1}0$ DF TEM images obtained from the squared region (i) marked in (a). (c) $g = \bar{1}010, g = \bar{2}110, g = \bar{1}\bar{1}20$, and $g = 1\bar{2}10$ DF TEM images obtained from the squared region (ii) in (a). Figure 2h, i in the main text are obtained by summing four different DF TEM images shown in (b) and (c), respectively to visualize the domain and domain wall contrasts. Note that the nematic boundaries visualized in $g = 2\bar{1}\bar{1}0$ DF TEM image in (b) and $g = 1\bar{2}10$ DF TEM image in (c) adopt different directions which are rotated 60° with respect to each other.

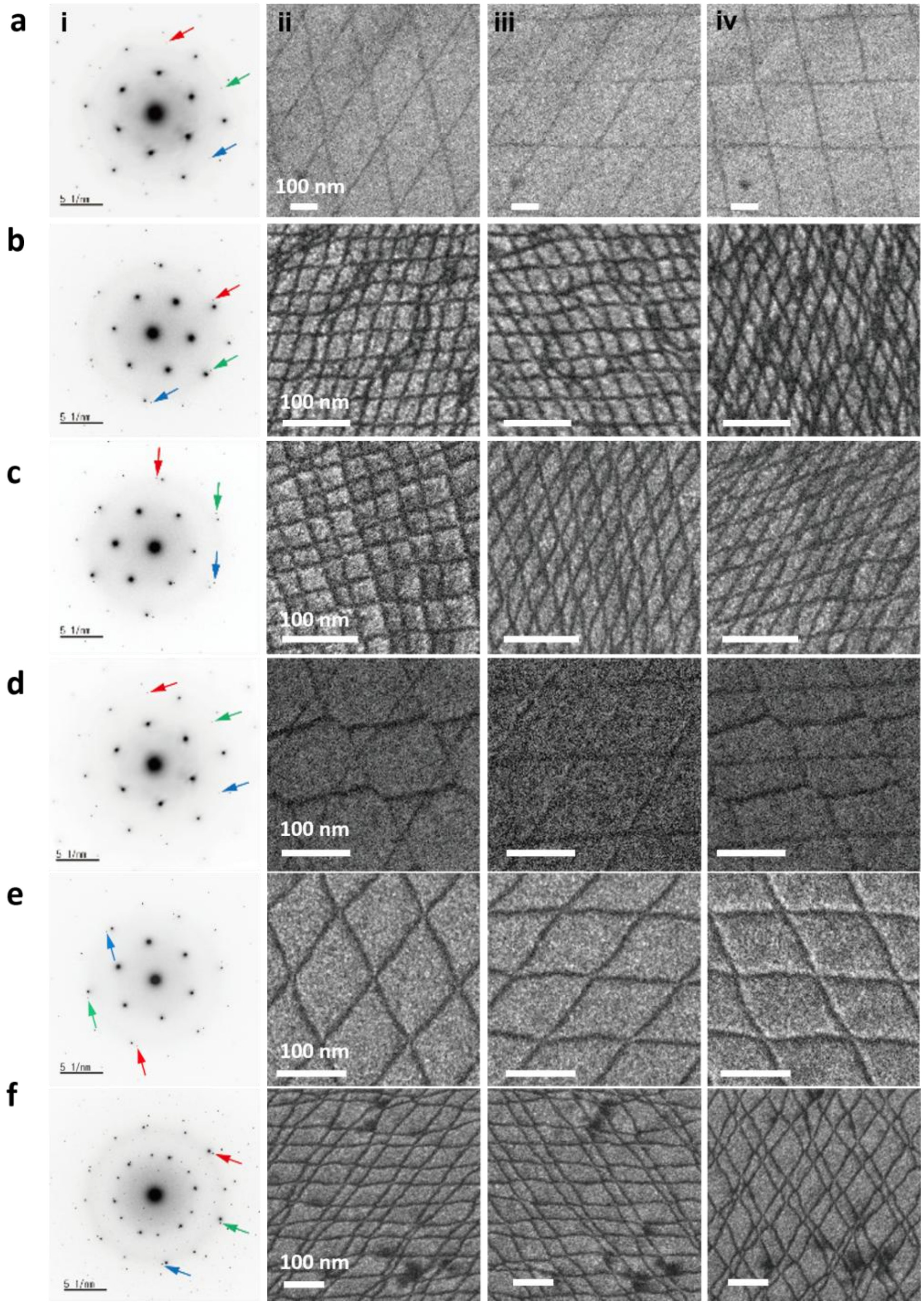

**Extended Data Fig. 4. Domain boundary contrast images obtained from TTG.** (a-f) SAED (panel i) and DF TEM images (panels ii-iv) utilizing the three sets of 2nd order Bragg peaks obtained from TTG with various twist angle combinations. Each of the Bragg peak used to obtain DF TEM images shown in panels ii, iii, iv are marked with red, green, and blue arrows in SAED. SAED and DF TEM images are obtained for the experimentally observed phases shown in the main text (Fig. 3): simple triangular domain lattice (a), simple triangular domain lattice bound by moiré-of-moiré boundaries (b), double colored triangular domain lattice (c), kagome-like domain lattice (d), colored triangular domain lattice (e), and hexagram domain lattice (f).

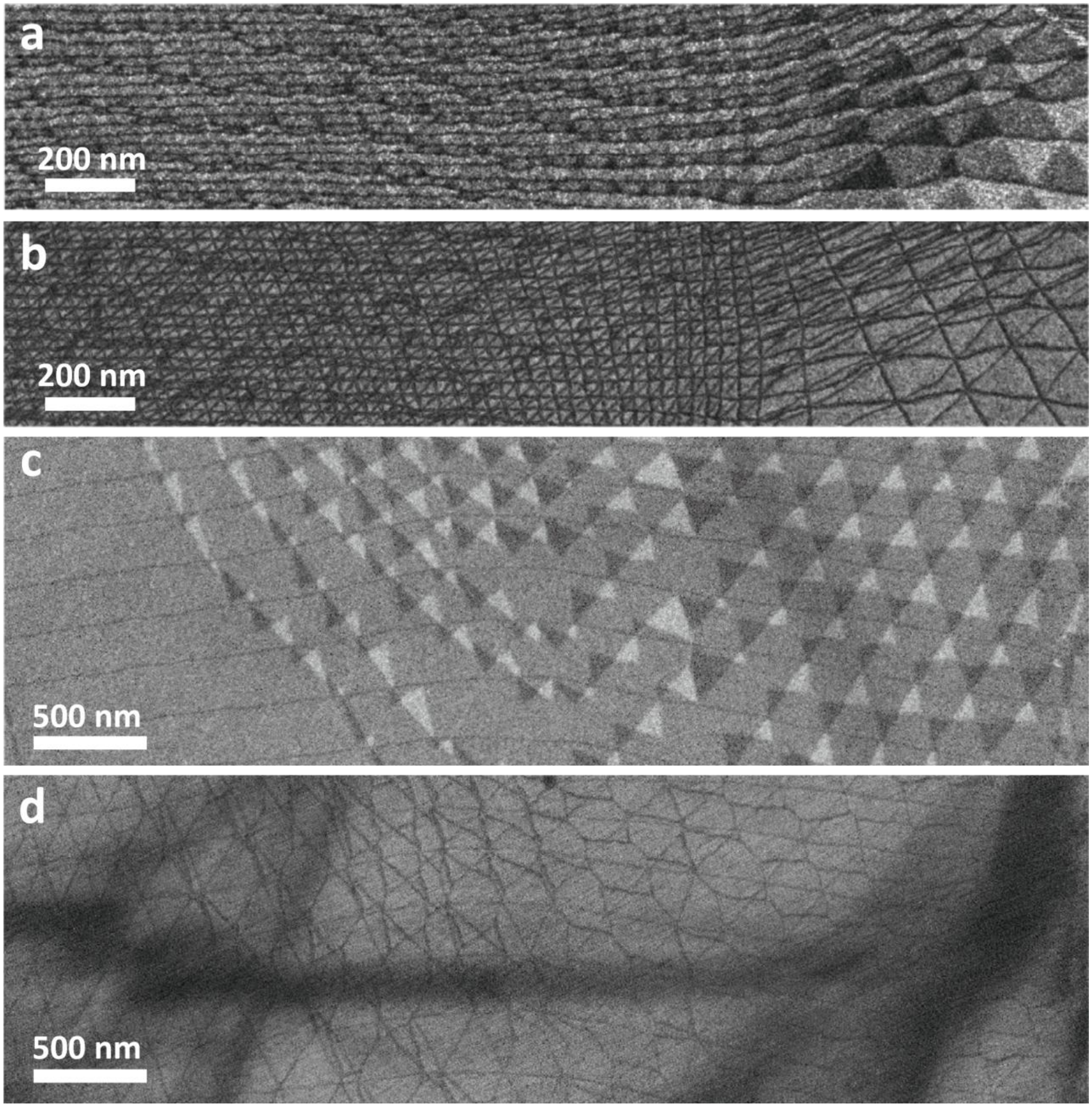

**Extended Data Fig. 5. Structural transition between different domain lattices.** (a-b) Domain contrast image ($g = 10\bar{1}0$ DF TEM image (a)) and the domain boundary contrast image (summation of $g = 2\bar{1}\bar{1}0, g = 11\bar{2}0$, and $g = \bar{1}2\bar{1}0$ DF TEM images (b)) obtained from the same region of the TTG specimen. Transition from the simple triangular domain lattice bound by moiré-of-moiré boundaries (left) to the double-colored triangular domain lattice near the commensurate condition of $2\theta_{12} + \theta_{23} = 0$ (center) to another locally commensurate colored triangular domain lattice (right). (c-d) Domain contrast image ($g = 10\bar{1}0$ DF TEM image (c)) and the domain boundary contrast image (summation of $g = 2\bar{1}\bar{1}0, g = 11\bar{2}0$, and $g = \bar{1}2\bar{1}0$ DF TEM images (d)) obtained from another region of the TTG specimen. Transition from the simple triangular domain lattice (left) to the kagome-like domain lattice (right) is shown to exhibit complex local minimum domain configuration in the middle.

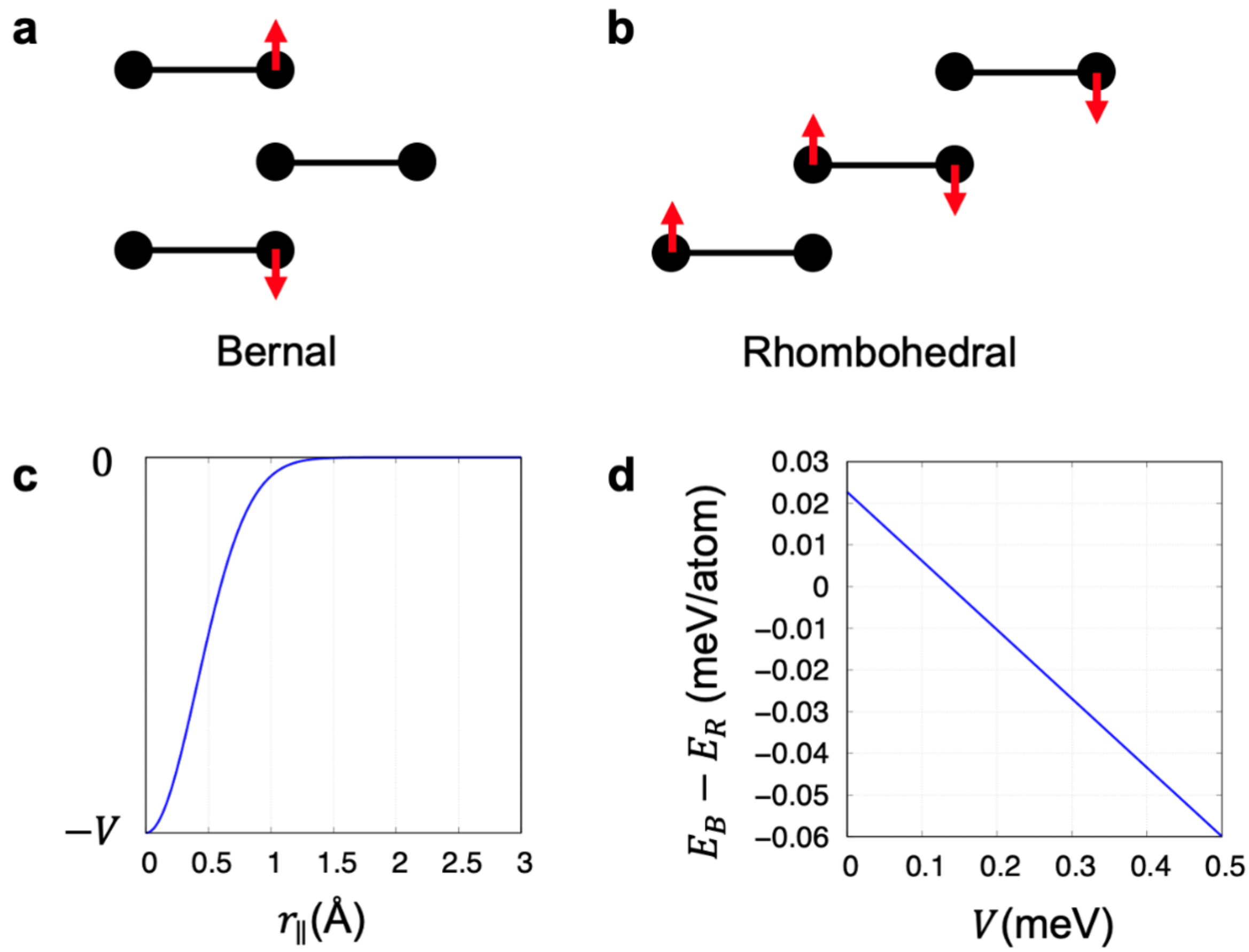

**Extended Data Fig. 6. Modified Kolmogorov-Crespi (KC) potential.** Spurious out-of-plane displacement of (a) Bernal- and (b) rhombohedral-stacked trilayer graphene. Red arrows indicate tiny but finite displacements around 1.4 x$10^{-3}$ Angstrom. (c) Gaussian-type the next-nearest neighbor interlayer interaction potential that depends only on the plane-projected distance $r_{\parallel}$ between carbon atoms on top and bottom layers. (d) The energy difference between Bernal ($E_B$) and rhombohedral ($E_R$) trilayer graphene as a function of control parameter $V$. Without $V$, $E_B$ is larger than $E_R$ with KC potential. With $V$ > 0.15 meV, the Bernal stacking become to be stabler than rhombohedral one.

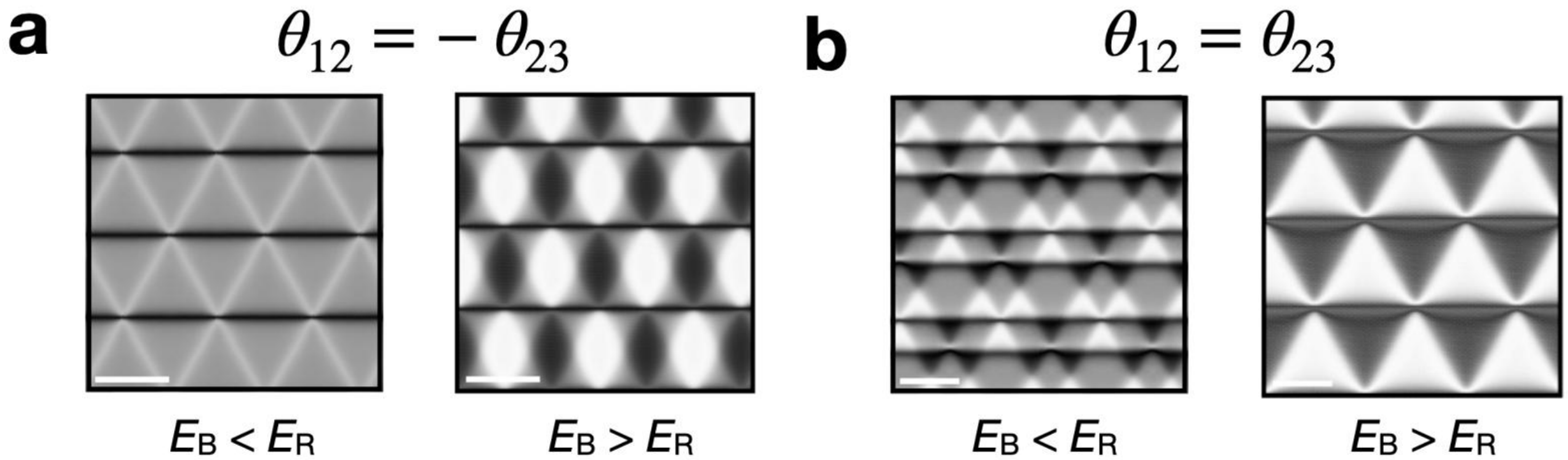

**Extended Data Fig. 7. Comparison of relaxed structures with different total energy configurations. a**. Simulated TEM images for alternate stacking case of $\theta_{12} + \theta_{23} = 0$, corresponding to the panel (i) in Fig. 3. Grey color corresponds to Bernal stacking and black and white colors to rhombohedral stacking. The scale bar is 100 nm. The left panel is a moiré domain pattern in case of Bernal stacking is the ground state while the right panel shows the fully relaxed domain pattern in case of rhombohedral stacking has a lower total energy than Bernal stacking. **b**. Simulated TEM images for helical stacking case of $\theta_{12} - \theta_{23} = 0$, corresponding to the panel (vi). The left and right panels are obtained with same conditions used in **a** and used same color scheme.

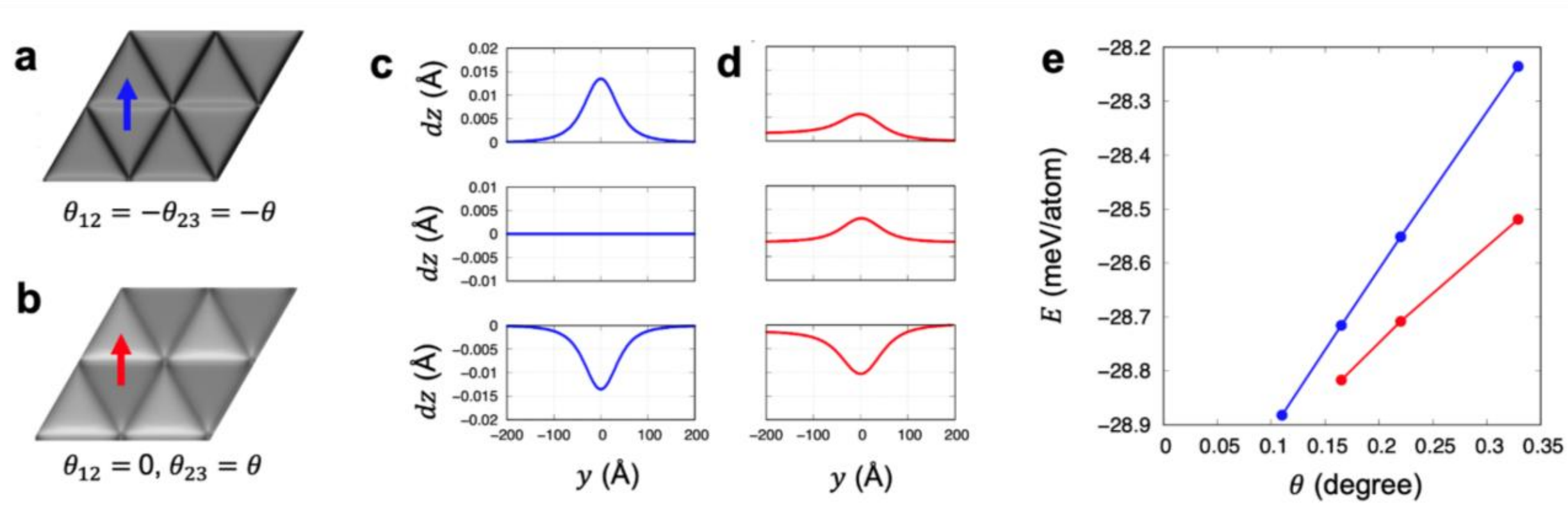

**Extended Data Fig. 8. Structures and energies of domain wall in TTG.** Stacking domain structures of TTG for (a) $\theta_{12} = -\theta_{23} = \theta$ and (b) $\theta_{12} = 0$, $\theta_{23} = \theta$. The height variation $dz$ along the blue (Bernal-Bernal boundary) and red arrows (Bernal-rhombohedral boundary) are drawn in (c) and (d), respectively. In (c) and (d), from the top to bottom panels, line profiles of $dz$ for the top, middle and bottom layers, respectively. (e) Energies ($E$) of TTG as a function of $\theta$ for the case (a) and (b) are plotted with filled blue and red circles, respectively. From the slopes of $E(\theta)$, the energies of domain walls are estimated to be 137 and 84 meV/Å, respectively.